\documentclass[lettersize,journal]{IEEEtran}
\usepackage{amsmath,amsfonts}
\usepackage{algorithmic}[line numbering]
\usepackage{algorithm}
\usepackage{array}
\usepackage[caption=false,font=normalsize,labelfont=sf,textfont=sf]{subfig}
\usepackage{textcomp}
\usepackage{stfloats}
\usepackage{url}
\usepackage{verbatim}
\usepackage{graphicx}
\def\BibTeX{{\rm B\kern-.05em{\sc i\kern-.025em b}\kern-.08em
    T\kern-.1667em\lower.7ex\hbox{E}\kern-.125emX}}
\usepackage{balance}
\usepackage{url}
\usepackage{verbatim}
\usepackage{graphicx}
\usepackage{scalerel}
\usepackage{diagbox}
\usepackage{tikz}
\usetikzlibrary{svg.path}
\usepackage{cite}
\usepackage[export]{adjustbox}
\usepackage{cuted}
\usepackage{amsmath,amssymb,amsfonts}
\usepackage{caption}
\usepackage{multirow}
\usepackage{float}
\usepackage{soul}

\usepackage{multicol}
\usepackage{makecell}
\usepackage{colortbl}
\usepackage{pgfplots}
\usepackage{lipsum}
\DeclareCaptionLabelSeparator{periodspace}{.\quad}
\pgfplotsset{compat=1.18}
\usepackage{balance}

\definecolor{orcidlogocol}{HTML}{A6CE39}
\tikzset{
 orcidlogo/.pic={
 \fill[orcidlogocol] svg{M256,128c0,70.7-57.3,128-128,128C57.3,256,0,198.7,0,128C0,57.3,57.3,0,128,0C198.7,0,256,57.3,256,128z};
 \fill[white] svg{M86.3,186.2H70.9V79.1h15.4v48.4V186.2z}
 svg{M108.9,79.1h41.6c39.6,0,57,28.3,57,53.6c0,27.5-21.5,53.6-56.8,53.6h-41.8V79.1z M124.3,172.4h24.5c34.9,0,42.9-26.5,42.9-39.7c0-21.5-13.7-39.7-43.7-39.7h-23.7V172.4z}
 svg{M88.7,56.8c0,5.5-4.5,10.1-10.1,10.1c-5.6,0-10.1-4.6-10.1-10.1c0-5.6,4.5-10.1,10.1-10.1C84.2,46.7,88.7,51.3,88.7,56.8z};}}
\newcommand\orcidicon[1]{\href{https://orcid.org/#1}{\mbox{\scalerel*{
\begin{tikzpicture}[yscale=-1,transform shape]
\pic{orcidlogo};
\end{tikzpicture}
}{|}}}}
\usepackage{hyperref}
\begin{document}

\title{
 $ULG_{SS}$: A Strategy to construct a Library of  Universal Logic Gates for $N$-variable Boolean Logic beyond NAND and NOR
}

\author{Aadarsh G. Goenka\orcidicon{0000-0002-7229-1471}, Shyamali Mitra\orcidicon{0000-0001-6502-9056}, Mrinal K. Naskar, Nibaran Das, \textit{Senior Member, IEEE}\orcidicon{0000-0002-2426-9915}

\thanks{A. G. Goenka (e-mail: aggtur11@gmail.com ) and N. Das (e-mail: nibaran@ieee.org) are with the Dept. of CSE, Jadavpur University, India }
\thanks{S. Mitra is with the Dept. of IEE , Jadavpur University, India (e-mail: shyamalimitra.iee@jadavpuruniversity.in)}
\thanks{M. K. Naskar is with the Dept. of ETCE , Jadavpur University, India (e-mail: mrinaletce@gmail.in)}
 }

\maketitle

\begin{abstract}

In literature, NAND and NOR are two logic gates that display functional completeness, hence regarded as Universal gates. So, the present effort is focused on exploring a library of universal  gates in binary that are still unexplored in literature along with a broad and systematic approach to classify the logic connectives. The  study shows that the number of Universal Gates in any logic system grows exponentially with the number of input variables $N$. It is revealed that there are $56$ Universal gates in binary for $N=3$. It is shown that the ratio of the count of Universal gates to the total number of Logic gates is $\approx $ $\frac{1}{4}$ or 0.25. Adding constants $0,1$ allow for the creation of $4$ additional (for $N=2$) and $169$ additional Universal Gates (for $N=3$). In this article, the mathematical and logical underpinnings of the concept of  universal logic gates are presented, along with a search strategy $ULG_{SS}$ exploring  multiple paths  leading to their identification. A fast-track approach has been introduced that uses the hexadecimal representation of a logic gate to quickly ascertain its attribute.

\end{abstract}

\section{Introduction}

Logic gates\cite{803116,Liu2021-ih,631214,9951401,9762783,9864039,Noiri2022,4468044} are the fundamental building blocks of any digital system \cite{AFPM,mitra2023low,GK}. An algebraic formula can be used to describe the operation of each gate, where each gate has its own unique graphic symbol. In Binary, computation is carried out by a lot of complex functions and logic gates \cite{WJ}. Logic gates are categorized into three different types depending on their properties; Basic gates, Universal Gates and Non-Universal Gates. Basic gates (AND, OR and NOT gates) are those that can form any logic function by combining some or all of these gates to make a functionally complete set \cite{AM}. Universal Gate(s) can construct any logic function without utilizing any other logic gate in the process; as a result, they may form a \textit{Functionally Complete Set} \cite{EP} themselves. Non-Universal gates, on the other hand, cannot construct all logic functions and must utilize other logic gates in the process. 

At present, there are two known Universal logic gates in binary: NAND and NOR gates, which indicate that it is possible to create any Boolean expression using only NOR or NAND gates. There are no more Universal Gates documented in literature. Aside from the fact that a Universal Gate should be able to produce AND, OR, and NOT gates, it is  unknown how to determine whether a  gate is Universal or not. This research unfolds several characteristics that may help to deduce the class of a logic connective using a broad and systematic approach. This leads to the identification of a library  of universal gates in the Binary Logic System. In order to illustrate the universality possessed by an $N$-variable logic, a number of  deductions are introduced. In addition to this, a discussion on how the logic of a Non-Universal logic changes when a full set consisting of logic constants $0$ and $1$  is included. A systematic approach in determining the nature of a given logic gate is introduced using some observations that explain the universality of a given logic. A mathematical expression to determine the number of universal logic gates is also presented to reveal that $R_N$, which is the ratio of universal logic gate count to all possible logic gates in the $N$-variable logic, increases linearly for $N=3$, and saturates to $\frac{1}{4}$ of the total number of gates for higher values of $N$. Further, the applicability of the logic gates has been shown in applications like quantum dot cellular automata (QCA) to design full adder circuits to reduce the overall hardware cost to a great extent. The contribution in the proposed work can be summarized as:

\vspace{2mm}

\begin{itemize}
\item Creating a library of universal logic gates in binary and showing their applicability in  certain applications.
\item Revisiting  the  conditions of universality of logic gates as well as establishing some new criteria for better analysis of the class of a logic with the proposed algorithms.
\item Deriving an expression to determine the number of universal gates in Boolean Logic with $N$-variables.
\item Behaviour of Multiplexer equivalent circuits consisting of a set \{0,1,G\}, where G is a logic gate in Binary.
\item A fast-track methodology for determining the property of the logic gate based on its hexadecimal encoding.
\end{itemize}
\begin{figure}[!t]
\centering
\includegraphics[width=0.47\textwidth]{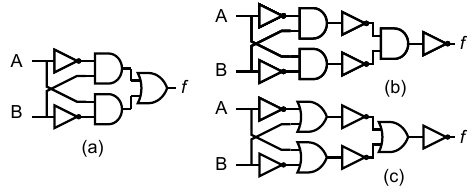}
\caption{Implementation of an expression $f=A\overline{B}+\overline{A}B$ using (a) AND-OR-NOT logic (b) AND-NOT logic (c) OR-NOT logic}
\label{fig:Complete_set}
\end{figure}

\vspace{-1mm}

\section{Universal Gates in Binary}

A functionally complete set of logic connectives or Boolean operators is one that, when members of the set are combined to form a Boolean expression they can express every possible Boolean logic. Every possible logic function can be realised as a network of gates specified in the set, which is referred to as functionally complete \cite{AC}. Logic Gates which are part of Functionally Complete set of cardinality $1$ are considered Universal. NAND and NOR are the examples of two singleton sets. In essence, all logic functions can be built using  binary NOR or NAND gates alone. If we cascade an AND gate and a NOT gate, we have a NAND gate, which is a complete set of logic similar to AND and NOT. To make a NOR gate, we would cascade an OR gate and a NOT gate. However, \{AND, OR\} alone do not create a complete set. If NOT gate is included in the set, the set behaves a complete set. The combination of these  Gates allows us to derive a wide variety of different Boolean operations and gates. However, the NAND and NOR gates are considered minimal sets due to their completeness; they can be used alone or in combination to build a wide variety of different logic circuits. The inference regarding a logic gate being universal can be drawn using the following properties:

\vspace{2mm}

 \textit{\textbf{Property 1: } A logic connective or a set of logic connectives is said to be functionally complete, if all possible logic functions are equivalent to Boolean Expressions involving only that gate or gates in the set.}
\par A set of Boolean functions is functionally complete if it is possible to generate all logic expressions using the set. A set is functionally complete if it can be derived from another set that is also functionally complete, e.g., \{AND, OR, NOT\},\{AND, NOT\}, \{OR, NOT\}. As an example, it is possible to construct all Boolean functions using a combination of AND, OR and NOT gates. An example is shown in Fig. \ref{fig:Complete_set} (a),(b),(c) where a function $f=A\overline{B}+B\overline{A}$ is implemented using three complete sets each consisting of all elements from the set. Thus, all elements from a complete set must be considered to realize a function. However, if  \{AND,OR\} is considered a set, only a subset of the functions can be realized, hence the set becomes functionally incomplete.

\textit{\textbf{Property 2: }A Universal gate of $N$-inputs can create all logic functions without using any other gate in the process.}

\vspace{1mm}

 Consider a Boolean expression $f$=$A\overline{B}+\overline{A}B$. This can be formed using NAND gates explicitly without involving any other gate in the process. 
The modified expression $f$=$\overline{\overline{A\overline{(AB)}}. \overline{B\overline{(AB)}}}$ allows us to implement the logic function using only NAND gates. In applications, this is beneficial because $NAND$ and $NOR$ gates are less expensive and simpler to fabricate. Furthermore, these gates are fundamental building blocks of all IC logic families.

\textit{\textbf{Property 3: }A logic is considered to be Non-Universal if it cannot produce a functionally complete set with cardinality $1$.}


Let us consider three sets with cardinality $1$; \{AND\}, \{OR\}, \{NOT\} which are functionally incomplete sets. From Table \ref{tab:2inputUniversalGate}, it is observed that, AND, OR, NOT gates can respectively implement only $3$, $3$ and $4$ logic functions without using logic constants $0,1$. Including the logic constants increase the number of implemented functions to $5$, $5$, and $6$ respectively. Hence, they cannot be considered as Universal gates.
\vspace{-1mm}
\subsection{Universal logic gates in Binary for N=2}
Logic gates, can be of one input ($A$) or two input variables ($A$ and $B$) and a single binary output variable (say, $Y$).
In Binary, there are $2^4=16$ possible Logic constructs. These range from $0000$ to $1111$ when $11$, $10$, $01$, and $00$ are given as inputs respectively to the two input lines of the gates. Thus, they can be numbered from $0$ to $15$ in decimal or $0$ to $F$ in hexadecimal. Table \ref{tab:2inputUniversalGate} shows each such logic and the number of  functions implemented by each logic. It is observed that only NOR (\#1) and NAND (\#7), are capable of implementing all $16$  logic functions and are thus regarded as Universal gates.

\begin{table}[H]
    \centering
    \caption{\# of   functions realized by a  gate with and without logic constants $0$,$1$ for $N=2$. Numbers in bold correspond to implementations of all logic combinations and hence considered Universal.}
   \resizebox{0.47\textwidth}{!}{\begin{tabular}{|c|c|c|c|}
    \hline
  \multirow{3}{*}{ \textbf{Gate \#}} & \multirow{3}{*}{\textbf{Logic function}} & \multicolumn{2}{c|}{\multirow{2}{*}{\textbf{\# of logic Implemented}}}                                                                                                                                           \\
                         &                        & \multicolumn{2}{c|}{}                                                                                                                                                                    \\ \cline{3-4} 
                         &                        & \multicolumn{1}{l|}{\begin{tabular}[c]{@{}l@{}}\textbf{Without  logic constants } \end{tabular}} & \multicolumn{1}{l|}{\begin{tabular}[c]{@{}l@{}}\textbf{With  logic constants} \end{tabular}} \\ \hline
    0 & F(A,B)=0 & 1 & 1 \\\hline
    1 & $\overline{A+B}$ & \textbf{16} & \textbf{16} \\\hline
    2 & $\overline{A}B$ & 6 & \textbf{16}   \\\hline
    3 & $\overline{A}$ & 4 & 6 \\\hline
    4 & $A\overline{B}$ & 6 & \textbf{16} \\\hline
    5 & $\overline{B}$ & 4 & 6 \\\hline
    6 & A xor B & 4 & 8 \\\hline
    7 & $\overline{AB}$ & \textbf{16} & \textbf{16} \\\hline
    8 & AB & 3 & 5 \\\hline
    9 & A xnor B & 4 & 8 \\\hline
    A/10 & B & 2 & 4 \\\hline
    B/11 & $\overline{A}+B$ & 6 & \textbf{16} \\\hline
    C/12 & A & 2 & 4 \\\hline
    D/13 & $A+\overline{B}$ & 6 & \textbf{16} \\\hline
    E/14 & A+B & 3 & 5 \\\hline
    F/15 & F(A,B)=1 & 1 & 1 \\\hline
    \end{tabular}} 
    \label{tab:2inputUniversalGate}
\end{table}

\begin{figure}[!t]
\small
\centering
\begin{tikzpicture}
\centering
\begin{axis}[
ybar interval=0.5,
width=7.2cm,
height=3.5cm,
legend style = { at = {(0.6,0.75)}},
xlabel=Gate \#,
ylabel=\# of Outcomes,
xmin=0,xmax=16,
ymin=0,ymax=17,
ytick={0,4,8,12,16},
grid,
]
\addplot [thin,black,fill=cyan] table {Data.dat};
\end{axis}
\end{tikzpicture}
\caption{Visualization of possible gates implemented for $N=2$}
\label{fig:visualization}
\end{figure}
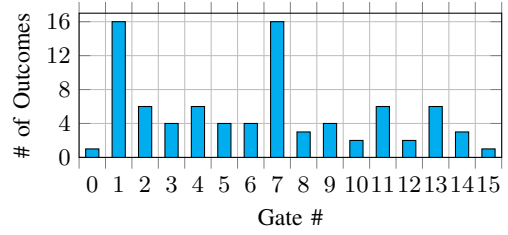
\vspace{-1mm}
\subsection{Universal logic in Binary for N=3}
In case of $N=3$, each gate has eight possible outcomes corresponding to $2^3$ input combinations ranging from $000$ to $111$. Thus, there are $2^{2^3}=256$ logic outcomes ranging from $00000000$ to $11111111$. The $256$ gates are numbered from $0$ to $255$ in decimal or $00$ to $FF$ in hexadecimal. Table \ref{tab:3input3LogicFunctions} shows the logic gates coded in hexadecimal from $00$ to $FF$  with the corresponding number of logic gates implemented by each. The gates that can implement all $256$ logic functions are Universal gates. It is found empirically that there are $56$ universal gates for $N=3$ fulfilling the criteria for the universal logic. A list of universal gates with the hexadecimal encoding are shown in Table \ref{tab:3inputUniversalGate}. Thus, dealing with higher number of variables increases the gate count proportionately.
\subsection{Observations and Analysis for Universal Gates}

In this section, we compile all observation notes, organize them, and analyze the data from the experimental outcomes.


\textit{\textbf{Observation 1:} The Universal Gates are all located in the first half of the total gate count and are all odd-indexed.} 

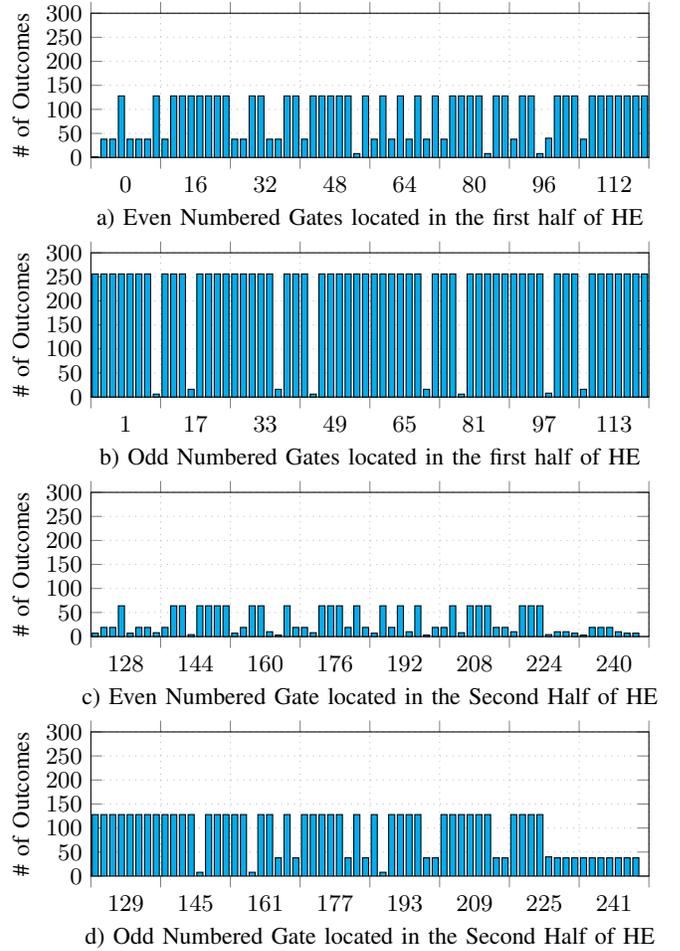
\begin{figure}[!t]
\begin{tikzpicture}
font=\small
\centering
\begin{axis}[
ybar interval=0.7,
width=9cm,
height=3.5cm,
legend style = { at = {(0.6,0.75)}},
xlabel=a) Even Numbered Gates located in the first half of HE,
ylabel=\# of  Outcomes,
xmin=0,xmax=128,
ymin=0,ymax=300,
ytick={0,50,100,150,200,250,300},
xtick={0, 16, ..., 128},
grid,
grid style={dotted}]
\addplot [thin,fill=cyan]table{Data3.dat};
\end{axis}
\end{tikzpicture}
\begin{tikzpicture}
font=\small
\centering
\begin{axis}[
ybar interval=0.7,
width=9cm,
height=3.5cm,
legend style = { at = {(0.6,0.75)}},
xlabel=b) Odd Numbered Gates located in the first half of HE,
ylabel=\# of  Outcomes,
xmin=1,xmax=129,
ymin=0,ymax=300,
ytick={0,50,100,150,200,250,300},
xtick={1, 17, ..., 129},
grid,
grid style={dotted}]
\addplot [thin,fill=cyan]table{Data3A.dat};
\end{axis}
\end{tikzpicture}

\begin{tikzpicture}
font=\small
\centering
\begin{axis}[
ybar interval=0.7,
width=9cm,
height=3.5cm,
legend style = { at = {(0.5,0.5)}},
xlabel=c) Even Numbered Gate located in the Second Half of HE,
ylabel=\# of  Outcomes,
xmin=128,xmax=256,
ymin=0,ymax=300,
ytick={0,50,100,150,200,250,300},
xtick={128, 144, ..., 256},
grid,
grid style={dotted}]
\addplot [thin,fill=cyan] table{Data3.dat};
\end{axis}
\end{tikzpicture}

\begin{tikzpicture}
font=\small
\centering
\begin{axis}[
ybar interval=0.7,
width=9cm,
height=3.5cm,
legend style = { at = {(0.6,0.75)}},
xlabel=d) Odd Numbered Gate located in the Second Half of HE,
ylabel=\# of  Outcomes,
xmin=129,xmax=257,
ymin=0,ymax=300,
ytick={0,50,100,150,200,250,300},
xtick={129, 145, ..., 257},
grid,
grid style={dotted}]
\addplot [thin,fill=cyan] table{Data3A.dat};
\end{axis}
\end{tikzpicture}
\caption{Visualization of the entire logic in a $2-D$ plane for $N=3$. The height of the bar graph denotes the number of different logic functions implemented by a given logic. Fig. \ref{graph1}b) shows that all universal logic gates (implementing all $256$ logic connectives) are entirely clustered in the first half of the hexadecimal encoding (HE) and that they are all odd numbered.}
\label{graph1}
\end{figure}

It may be deduced from Fig. \ref{graph1}b) that all the Universal Gates  always lie in the first half of the total gate count when arranged in ascending order of their hexadecimal encoding i.e., output will be $0$ when all inputs are $1$, and that they will always be odd numbered, regardless of the number of inputs that a binary gate can accept (i.e., output will be $1$ when all inputs are $0$). If a gate  fails to meet any one of these two conditions, it is said to be Non-Universal. A gate that satisfies both of these criteria can be either Universal or non-Universal depending on fulfillment of further criteria explained below.

\begin{table}[H]
\caption{List of $56$ Universal Gates with their H-E for $N=3$}
\centering
\resizebox{0.32\textwidth}{!}{\begin{tabular}{|c|c|c|c|c|c|c|c|}
\hline
01 & 03 & 05 & 07 & 09 & 0B & 0D & 11\\\hline
13 & 15 & 19 & 1B & 1D & 1F & 21 & 23\\\hline
25 & 27 & 29 & 2D & 2F & 31 & 35 & 37\\\hline
39 & 3B & 3D & 3F & 41 & 43 & 45 & 47\\\hline
49 & 4B & 4F & 51 & 53 & 57 & 59 & 5B\\\hline 
5D & 5F & 61 & 63 & 65 & 67 & 6B & 6D\\\hline
6F & 73 & 75 & 77 & 79 & 7B & 7D & 7F\\\hline
\end{tabular}}

\label{tab:3inputUniversalGate}
\end{table}
 
\textit{\textbf{Observation 2:} It is implicit that for $N=2$, out of $16$ Logic Gates, $4$  satisfy both the criteria stated in $Observation 1$ and out of $4$ gates, $2$ are Universal. Similarly, for $N=3$, out of $256$  gates, $64$ satisfy both criteria, and only $56$ are Universal.}

\begin{table}[H]
\caption{Truth Table of Non-Universal gates with $N=3$}
\centering
\resizebox{0.84\linewidth}{!}{
\begin{tabular}{|c|c|c|c|c|c|c|c|c|c|c|}
\hline
\textbf{A} & \textbf{B} & \textbf{C} & \textbf{\#0F} & \textbf{\#17} & \textbf{\#2B} & \textbf{\#33} & \textbf{\#4D} & \textbf{\#55} & \textbf{\#69} & \textbf{\#71} \\\hline
0 & 0 & 0 & 1 & 1 & 1 & 1 & 1 & 1 & 1 & 1 \\\hline
0 & 0 & 1 & 1 & 1 & 1 & 1 & 0 & 0 & 0 & 0 \\\hline
0 & 1 & 0 & 1 & 1 & 0 & 0 & 1 & 1 & 0 & 0 \\\hline
0 & 1 & 1 & 1 & 0 & 1 & 0 & 1 & 0 & 1 & 0 \\\hline
1 & 0 & 0 & 0 & 1 & 0 & 1 & 0 & 1 & 0 & 1 \\\hline
1 & 0 & 1 & 0 & 0 & 1 & 1 & 0 & 0 & 1 & 1 \\\hline
1 & 1 & 0 & 0 & 0 & 0 & 0 & 1 & 1 & 1 & 1 \\\hline
1 & 1 & 1 & 0 & 0 & 0 & 0 & 0 & 0 & 0 & 0 \\\hline
\end{tabular}
}
\label{tab:3inputNonUniversalGate}
\end{table}

\begin{table}[H]
\caption{Truth Tables of gate $3$ and $5$}
\centering
\begin{tabular}{|c|c|c|c|}
\hline
\textbf{A} & \textbf{B} & \textbf{\#3} & \textbf{\#5} \\\hline
0 & 0 & 1 & 1 \\\hline
0 & 1 & 1 & 0 \\\hline
1 & 0 & 0 & 1 \\\hline
1 & 1 & 0 & 0 \\\hline
\end{tabular}
\label{tab:2inputGate3and5}
\end{table}


 Four gates located in the odd-numbered positions of hexadecimal encoding are  $1$, $3$, $5$, and $7$. But Gate $3$ and Gate $5$ are non-universal. The reason can be explained with the help of truth tables shown in Table \ref{tab:2inputGate3and5}. Thus, gates $1$ and $3$ satisfy both the criteria stated in $Observation 1$. Similarly, Gates $F, 17, 2B, 33, 4D, 55, 69$ and $71$ in $N=3$ satisfy both the criteria and are still Non-Universal. The truth tables of these non-universal gates are given in Table \ref{tab:3inputNonUniversalGate}.
It can be observed that they all have equal number of $1$s and $0$s in their output. In addition, the first and second halves of the output of the gates have inversely mirrored arrangements of $0$s and $1$s. Therefore, each of them must satisfy,

\begin{equation}
\overline{F(A,B)}=F(\overline{A},\overline{B})
\end{equation} 
 \vspace{-2mm}
\begin{equation}
\overline{F(A,B,C)}=F(\overline{A},\overline{B},\overline{C})
\end{equation}

\subsection{Data Interpretation using the logic analysis of Gates}

To be expressively complete, any set of $n$-valued truth functional connectives must satisfy all of the conditions outlined in section II. This posits that a set of connectives is expressively complete if none of the highlighted properties that an arbitrary connective may possess are satisfied. It is stated in terms of the three conditions listed below.

\textit{\textbf{Condition 1:} A logic gate $G$ which is in the class of zero-preserving boolean functions $T0$ (i.e., $f \in T0$ if it satisfies f(0,0,..,0)=0) is always Non-Universal.}

\vspace{1mm}

\textit{\textbf{Proof:}} To express that the proposition is true, let us consider a logic gate $G$ which can be expressed as, $f(0,0,0...0)=0$. In other way, for the gate to create $f1(0,0,0...0) = 1$ at least one of the inputs must be $1$. This requires an extra NOT Gate in conjunction with it or any gate that manifest itself in the form of $f1$ (i.e., it produces a $1$ when all the inputs are $0$). Thus, an additional gate is needed to create $f1$. Therefore, $f1$ cannot be produced with $G$ alone; hence $f$ is non-universal.

\vspace{2mm}

\textit{\textbf{Condition 2:} A logic gate $G$ which is in the class of one-preserving boolean functions $T1$ (i.e., $f \in T1$ if it satisfies f(1,1,..,1)=1) is always Non-Universal.}

\vspace{2mm}

\textit{\textbf{Proof:}} To express that the proposition is true, let us consider a logic gate $G$ where, $f(1,1,1...1)=1$. Therefore, for the gate to create $f1(1,1,1...1) = 0$, at least one of the inputs must be $0$. This requires a NOT Gate or any gate that is itself in the form of $f1$ (i.e. it gives 0 when all the inputs are 1). Thus, an additional gate is needed to create $f1$. Therefore, $f1$ cannot be produced with Gate $f$ alone; hence $f$ is Non-Universal.

\vspace{2mm}

Consider the gate $F8$ in a $3$ variable Boolean logic given by, $f=(A+BC)$, which does not satisfy any of the conditions of universality i.e., $f(0,0,0)=1$ (non-zero preserving) and $f(1,1,1)=0$ (non-one preserving). $F8$ is capable of producing the following Logic Gates: $AA, CC, EA, EC, EE, F0, F8, FA, FC$, and $FE$. These gates meet the criteria of a $0$-preserving function, $f(0,0,0)=0$, and a $1$-preserving function, $f(1,1,1)=1$.

\vspace{2mm}

\textit{\textbf{Condition 3:} A logic gate $F$ belonging to a class of self-dual Boolean functions $S$, $f \in S$ and satisfies $\overline{f(A,B,C...)} = f(\overline{A},\overline{B},\overline{C}...)$ is always Non-Universal.}

\vspace{2mm}

 Consider Gate $2B$ with $f$=($\overline{A+B}+\overline{A}C+\overline{B}C$), which satisfies the equation $\overline{F(A,B,C)} = F(\overline{A},\overline{B},\overline{C})$. The truth table is given in Table \ref{tab:3inputNonUniversalGate} which shows the function is self-dual, hence considered as non-universal. The possible Logic Gates that can be created by it are 0F, 17, 33, 4D, 55, 69, 71, 8E, 96, AA, B2, CC, D4, E8 and F0. All of the gates satisfy the equation $\overline{F(A,B,C)} = F(\overline{A},\overline{B},\overline{C})$. Since these are all generated using a non-universal logic, the implemented gates become non-universal by virtue.

\vspace{2mm}

\textit{\textbf{Condition 4}: Functions that neither preserve $0$ nor $1$ nor do they belong to the class of self-dual functions are said to be Universal. }

The proposed set of connectives is functionally complete as neither of the conditions $f(0,0,0...0)=0$, nor $f(1,1,1...1)= 1$, nor $\overline{f(A,B,C...)}= f(\overline{A},\overline{B},\overline{C}...)$ are satisfied. A logic which is functionally complete does not satisfy the condition for self-duality such that, there occurs a row in the truth table $\{y_1,y_2... ,y_n\}$ so, $f(y_1,y_2,... ,y_n) =f(-y_1,-y_2... ,-y_n)$.

\vspace{-2mm}

\subsection{A systematic approach for classifying Boolean Logic}

In this section, a systematic method for determining the class of a logic connective is presented. In order for the proposed set of connectives to be functionally complete, each of the theorems listed below must be satisfied by a connective in the set. We begin by outlining these properties.

\vspace{2mm}

\textit{\textbf{Theorem 1:} A Logic Gate which is a part of a Functionally Complete Set of cardinality 1 is a Universal Gate.}

\textit{\textbf{Proof:}} 
If a set consists of only one unique element, say, \{NAND\} and \{NOR\}, then this is regarded as functionally complete set with cardinality $1$. The interpretation of which is, that the constituent element can form any logic without requiring any other gate in the process.

\textit{\textbf{Theorem 2:} A logic gate of N-inputs that can create any other Universal Gate is also a Universal Gate. }

\textit{\textbf{Proof:}} Take into consideration a gate $G$ that has the ability to produce another gate $G1$. If this is the case, the set of gates that is produced by $G$ will incorporate the logic that is produced by $G1$. If $G1$ is a Universal Gate, then the set of logic connectives that it generates will include $2^{2^N}$ possible gates without the need for any additional gates to be used at any point in the process. As $G1$ is contained within $G$, it is deducible that $G$ is a Universal logic. Thus, it is possible to create all logic functions by utilising $G$ on its own. Take, for illustration, a gate that generates a NOR gate, which is a universal logic. NAND gate can solely generate a NOR logic and hence considered to be universal.
\par \textit{\textbf{Corallary 1:} A Non-Universal Gate can generate only Non-Universal logic for all possible inputs.}
\par The correctness of this statement can be stated using \textit{\textbf{Theorem 2}}, that recognise a general transformation of the proposition. For the 3-Input Non Universal Gate $2B$, the possible logic gates that it can implement are: $0F$, $17$, $2B$, $33$, $4D$, $55$, $69$, $71$, $8E$, $96$, $AA$, $B2$, $CC$, $D4$, $E8$ and $F0$. These instances may serve to infer that all of these gates are Non-Universal as they are expressed by a non-universal logic that always excludes a universal set. The algorithm for finding a universal logic is shown in \textbf{\textit{Algorithm 1}}.

\begin{algorithm}[H]
\begin{algorithmic}[1]
\small
\caption{\textit{Search for Universal Logic Gate}} 
\label{Algorithm:Uni}
\STATE \textbf{Input:} $TT$ \textit{{//Truth Table of a given logic with N inputs.}}
\\\textbf{Output: }Universal Gate, Non-Universal Gate.
\STATE \textbf{Variable Initialization:}
\\$N$ = Number of Inputs
\\$M=2^N$ \textit{{//No. of input logic combinations.}}
\\Initialize $TT[\;]$ with the Values from Truth Table of a function
\\$i \gets 1$
 \IF{$(TT[0]=1$ \&\& $TT[M-1]=0)$} 
 \WHILE{$i \leq \frac{M}{2}$}
 \IF{$TT[i]=TT[M-i-1]$}
 \RETURN {\textit{Universal Gate}}
\ELSE 
\STATE $i \gets i+1$
\ENDIF
\ENDWHILE
\ENDIF
\RETURN \textit{Non-Universal Gate}
\end{algorithmic}
 \end{algorithm} 

 \begin{table*}
 \centering
 \caption{Showing the number of   functions realized by each $256$ gates for $N=3$. Cells in yellow shows the realization of all functions in the set.}
 \resizebox{0.82\textwidth}{!}{
 \begin{tabular}{|c|c|c|c|c|c|c|c|c|c|c|c|c|c|c|c|}
 \hline
 \textbf{Gate \#} & \textbf{Output} & \textbf{Gate \#} & \textbf{Output} & \textbf{Gate \#} & \textbf{Output} & \textbf{Gate \#} & \textbf{Output} & \textbf{Gate \#} & \textbf{Output} & \textbf{Gate \#} & \textbf{Output} & \textbf{Gate \#} & \textbf{Output} & \textbf{Gate \#}  & \textbf{Output} \\ \hline
 00 & 1 & \cellcolor[HTML]{FFFFC7}01 & \cellcolor[HTML]{FFFFC7}256 & 02 & 38 & \cellcolor[HTML]{FFFFC7}03 & \cellcolor[HTML]{FFFFC7}256 & 04 & 38 & \cellcolor[HTML]{FFFFC7}05 & \cellcolor[HTML]{FFFFC7}256 & 06 & 128 & \cellcolor[HTML]{FFFFC7}07 & \cellcolor[HTML]{FFFFC7}256 \\ \hline
 08 & 38 & \cellcolor[HTML]{FFFFC7}09 & \cellcolor[HTML]{FFFFC7}256 & 0A & 38 & \cellcolor[HTML]{FFFFC7}0B & \cellcolor[HTML]{FFFFC7}256 & 0C & 38 & \cellcolor[HTML]{FFFFC7}0D & \cellcolor[HTML]{FFFFC7}256 & 0E & 128 & 0F & 6 \\ \hline
 10 & 38 & \cellcolor[HTML]{FFFFC7}11 & \cellcolor[HTML]{FFFFC7}256 & 12 & 128 & \cellcolor[HTML]{FFFFC7}13 & \cellcolor[HTML]{FFFFC7}256 & 14 & 128 & \cellcolor[HTML]{FFFFC7}15 & \cellcolor[HTML]{FFFFC7}256 & 16 & 128 & 17 & 16 \\ \hline
 18 & 128 & \cellcolor[HTML]{FFFFC7}19 & \cellcolor[HTML]{FFFFC7}256 & 1A & 128 & \cellcolor[HTML]{FFFFC7}1B & \cellcolor[HTML]{FFFFC7}256 & 1C & 128 & \cellcolor[HTML]{FFFFC7}1D & \cellcolor[HTML]{FFFFC7}256 & 1E & 128 & \cellcolor[HTML]{FFFFC7}1F & \cellcolor[HTML]{FFFFC7}256 \\ \hline
 20 & 38 & \cellcolor[HTML]{FFFFC7}21 & \cellcolor[HTML]{FFFFC7}256 & 22 & 38 & \cellcolor[HTML]{FFFFC7}23 & \cellcolor[HTML]{FFFFC7}256 & 24 & 128 & \cellcolor[HTML]{FFFFC7}25 & \cellcolor[HTML]{FFFFC7}256 & 26 & 128 & \cellcolor[HTML]{FFFFC7}27 & \cellcolor[HTML]{FFFFC7}256 \\ \hline
 28 & 38 & \cellcolor[HTML]{FFFFC7}29 & \cellcolor[HTML]{FFFFC7}256 & 2A & 38 & 2B & 16 & 2C & 128 & \cellcolor[HTML]{FFFFC7}2D & \cellcolor[HTML]{FFFFC7}256 & 2E & 128 & \cellcolor[HTML]{FFFFC7}2F & \cellcolor[HTML]{FFFFC7}256 \\ \hline
 30 & 38 & \cellcolor[HTML]{FFFFC7}31 & \cellcolor[HTML]{FFFFC7}256 & 32 & 128 & 33 & 6 & 34 & 128 & \cellcolor[HTML]{FFFFC7}35 & \cellcolor[HTML]{FFFFC7}256 & 36 & 128 & \cellcolor[HTML]{FFFFC7}37 & \cellcolor[HTML]{FFFFC7}256 \\ \hline
 38 & 128 & \cellcolor[HTML]{FFFFC7}39 & \cellcolor[HTML]{FFFFC7}256 & 3A & 128 & \cellcolor[HTML]{FFFFC7}3B & \cellcolor[HTML]{FFFFC7}256 & 3C & 8 & \cellcolor[HTML]{FFFFC7}3D & \cellcolor[HTML]{FFFFC7}256 & 3E & 128 & \cellcolor[HTML]{FFFFC7}3F & \cellcolor[HTML]{FFFFC7}256 \\ \hline
 40 & 38 & \cellcolor[HTML]{FFFFC7}41 & \cellcolor[HTML]{FFFFC7}256 & 42 & 128 & \cellcolor[HTML]{FFFFC7}43 & \cellcolor[HTML]{FFFFC7}256 & 44 & 38 & \cellcolor[HTML]{FFFFC7}45 & \cellcolor[HTML]{FFFFC7}256 & 46 & 128 & \cellcolor[HTML]{FFFFC7}47 & \cellcolor[HTML]{FFFFC7}256 \\ \hline
 48 & 38 & \cellcolor[HTML]{FFFFC7}49 & \cellcolor[HTML]{FFFFC7}256 & 4A & 128 & \cellcolor[HTML]{FFFFC7}4B & \cellcolor[HTML]{FFFFC7}256 & 4C & 38 & 4D & 16 & 4E & 128 & \cellcolor[HTML]{FFFFC7}4F & \cellcolor[HTML]{FFFFC7}256 \\ \hline
 50 & 38 & \cellcolor[HTML]{FFFFC7}51 & \cellcolor[HTML]{FFFFC7}256 & 52 & 128 & \cellcolor[HTML]{FFFFC7}53 & \cellcolor[HTML]{FFFFC7}256 & 54 & 128 & 55 & 6 & 56 & 128 & \cellcolor[HTML]{FFFFC7}57 & \cellcolor[HTML]{FFFFC7}256 \\ \hline
 58 & 128 & \cellcolor[HTML]{FFFFC7}59 & \cellcolor[HTML]{FFFFC7}256 & 5A & 8 & \cellcolor[HTML]{FFFFC7}5B & \cellcolor[HTML]{FFFFC7}256 & 5C & 128 & \cellcolor[HTML]{FFFFC7}5D & \cellcolor[HTML]{FFFFC7}256 & 5E & 128 & \cellcolor[HTML]{FFFFC7}5F & \cellcolor[HTML]{FFFFC7}256 \\ \hline
 60 & 38 & \cellcolor[HTML]{FFFFC7}61 & \cellcolor[HTML]{FFFFC7}256 & 62 & 128 & \cellcolor[HTML]{FFFFC7}63 & \cellcolor[HTML]{FFFFC7}256 & 64 & 128 & \cellcolor[HTML]{FFFFC7}65 & \cellcolor[HTML]{FFFFC7}256 & 66 & 8 & \cellcolor[HTML]{FFFFC7}67 & \cellcolor[HTML]{FFFFC7}256 \\ \hline
 68 & 40 & 69 & 8 & 6A & 128 & \cellcolor[HTML]{FFFFC7}6B & \cellcolor[HTML]{FFFFC7}256 & 6C & 128 & \cellcolor[HTML]{FFFFC7}6D & \cellcolor[HTML]{FFFFC7}256 & 6E & 128 & \cellcolor[HTML]{FFFFC7}6F & \cellcolor[HTML]{FFFFC7}256 \\ \hline
 70 & 38 & 71 & 16 & 72 & 128 & \cellcolor[HTML]{FFFFC7}73 & \cellcolor[HTML]{FFFFC7}256 & 74 & 128 & \cellcolor[HTML]{FFFFC7}75 & \cellcolor[HTML]{FFFFC7}256 & 76 & 128 & \cellcolor[HTML]{FFFFC7}77 & \cellcolor[HTML]{FFFFC7}256 \\ \hline
 78 & 128 & \cellcolor[HTML]{FFFFC7}79 & \cellcolor[HTML]{FFFFC7}256 & 7A & 128 & \cellcolor[HTML]{FFFFC7}7B & \cellcolor[HTML]{FFFFC7}256 & 7C & 128 & \cellcolor[HTML]{FFFFC7}7D & \cellcolor[HTML]{FFFFC7}256 & 7E & 128 & \cellcolor[HTML]{FFFFC7}7F & \cellcolor[HTML]{FFFFC7}256 \\ \hline
 80 & 7 & 81 & 128 & 82 & 19 & 83 & 128 & 84 & 19 & 85 & 128 & 86 & 64 & 87 & 128 \\ \hline
 88 & 7 & 89 & 128 & 8A & 19 & 8B & 128 & 8C & 19 & 8D & 128 & 8E & 8 & 8F & 128 \\ \hline
 90 & 19 & 91 & 128 & 92 & 64 & 93 & 128 & 94 & 64 & 95 & 128 & 96 & 4 & 97 & 128 \\ \hline
 98 & 64 & 99 & 8 & 9A & 64 & 9B & 128 & 9C & 64 & 9D & 128 & 9E & 64 & 9F & 128 \\ \hline
 A0 & 7 & A1 & 128 & A2 & 19 & A3 & 128 & A4 & 64 & A5 & 8 & A6 & 64 & A7 & 128 \\ \hline
 A8 & 10 & A9 & 128 & AA & 3 & AB & 38 & AC & 64 & AD & 128 & AE & 19 & AF & 38 \\ \hline
 B0 & 19 & B1 & 128 & B2 & 8 & B3 & 128 & B4 & 64 & B5 & 128 & B6 & 64 & B7 & 128 \\ \hline
 B8 & 64 & B9 & 128 & BA & 19 & BB & 38 & BC & 64 & BD & 128 & BE & 19 & BF & 38 \\ \hline
 C0 & 7 & C1 & 128 & C2 & 64 & C3 & 8 & C4 & 19 & C5 & 128 & C6 & 64 & C7 & 128 \\ \hline
 C8 & 10 & C9 & 128 & CA & 64 & CB & 128 & CC & 3 & CD & 38 & CE & 19 & CF & 38 \\ \hline
 D0 & 19 & D1 & 128 & D2 & 64 & D3 & 128 & D4 & 8 & D5 & 128 & D6 & 64 & D7 & 128 \\ \hline
 D8 & 64 & D9 & 128 & DA & 64 & DB & 128 & DC & 19 & DD & 38 & DE & 19 & DF & 38 \\ \hline
 E0 & 10 & E1 & 128 & E2 & 64 & E3 & 128 & E4 & 64 & E5 & 128 & E6 & 64 & E7 & 128 \\ \hline
 E8 & 4 & E9 & 40 & EA & 10 & EB & 38 & EC & 10 & ED & 38 & EE & 7 & EF & 38 \\ \hline
 F0 & 3 & F1 & 38 & F2 & 19 & F3 & 38 & F4 & 19 & F5 & 38 & F6 & 19 & F7 & 38 \\ \hline
 F8 & 10 & F9 & 38 & FA & 7 & FB & 38 & FC & 7 & FD & 38 & FE & 7 & FF & 1 \\ \hline
 \end{tabular}} \label{tab:3input3LogicFunctions}
 \end{table*}
\subsection{ Universal Gate count for N-variables}

Let, the total number of input combinations for a given logic with $N$ number of input variables is $I_N$, given by $I_N=2^N$. 
So, the total number of possible logic functions in an $N$ input binary system is $G_N=2^{I_N}$. 
The total number of input combinations for a particular gate which are unrestricted by \textbf{\textit{Condition 1}} and \textbf{\textit{Condition 2}} is given by $2^N-2$. The number of logic gates that are neither $0$-preserving nor $1$-preserving can be given by $G_S$ =$2^{2^N-2}$ = $\frac{2^{I_N}}{2^2}$ = $\frac{G_N}{4}$. The total number of input combinations for a particular gate that are unrestricted by all \textbf{\textit{Condition 1}}, \textbf{\textit{Condition 2}} and \textbf{\textit{Condition 3}} = $\frac{(2^N-2)}{2}$. The total number of logic connectives that satisfy the condition of self-duality but are neither $0$ nor $1$ preserving is given by, $2^{\frac{(2^N-2)}{2}}$ = $\sqrt{\frac{2^{I_N}}{2^2}}$ = $\sqrt{\frac{G_N}{4}}$. Therefore, the total number of universal gates $U_N$ for an $N$ input system can be stated as, 
\begin{equation}\footnotesize
 U_N = \frac{G_N}{4}-\sqrt{\frac{G_N}{4}}
\end{equation}

The ratio ($R_N$) of Universal Gate  count ($U_N$) to the total  gate  count ($G_N$)  in an $N$-variable logic system is given by,
\begin{equation}\footnotesize
 R_N = \frac{U_N}{G_N}=\frac{1}{4}-\frac{1}{\sqrt{4G_N}}
\end{equation}

Thus, from Eq. $3$ the Universal Gate count for $N=3$ is,

$I_3=2^3=8$, $G_3=2^{I_3}=256$, $U_3={\frac{256}{4}}-\sqrt{\frac{256}{4}}=\textbf{56}$.

\vspace{1mm}

Similarly, the  Universal gate count when $N=4$ is,
$I_4=2^4=16$, $G_4=2^{I_4}=65536$, $U_4=\frac{65536}{4}-\sqrt{\frac{65536}{4}}=\textbf{16256}$. 

\vspace{1mm}

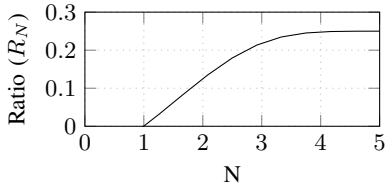
\begin{figure}[!t]
\small
\centering
\begin{tikzpicture}
\centering
\begin{axis}[width=5.5cm,height=3.1cm,
legend style = { at = {(0.6,0.75)}},
xlabel=N,
ylabel=Ratio ($R_N$),
xmin=0,xmax=5,
ymin=0,ymax=0.3,
xtick={0,1,2,3,4,5},
grid,
grid style={dotted}]
\addplot [thin] {0.25-(0.5/(2^(2^(x-1))))};
\end{axis}
\end{tikzpicture}
\caption{Ratio $R_N$ for different values of $N$}
\label{fig:Ratio_Rn}
\end{figure}

It is observed from Fig. \ref{fig:Ratio_Rn} that the ratio $R_N$ saturates to $0.25$, as the influence of the second term can be neglected for higher values of $N$, i.e., from $N>3$.

\section{Universal Logic Gates in Binary including logic constants 0,1}

In this section, the logic constants $0,1$ have been incorporated  to form a larger set of universal logic gates. A logic gate $G$ which is a part of \textit{Functionally Complete set} in the form $\{G,0,1\}$ is considered to be Universal, where, $1$ is the logic gate when $F(a_1,a_2,..,a_n)=1$ and $0$ is the logic gate which satisfies $F(a_1,a_2,..,a_n)=0$.
\subsection{Universal logic (N=2) including the logic constants 0, 1}

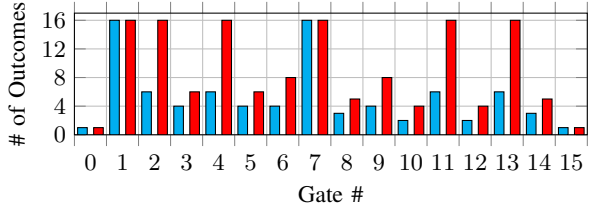
\begin{figure}[!t]
\small
\centering
\begin{tikzpicture}
\centering
\begin{axis}[
ybar interval=0.6,
width=8.4cm,
height=3.2cm,
legend style = { at = {(0.6,0.75)}},
xlabel=Gate \#,
ylabel= \# of Outcomes,
xmin=0,xmax=16,
ymin=0,ymax=17,
ytick={0,4,8,12,16},
grid,
]
\addplot [thin,fill=cyan] table {Data.dat};
\addplot [thin,fill=red] table {Data2.dat};
\end{axis}
\end{tikzpicture}
\caption{Increase in the number of logic realized by a gate (shown in red color) for $N=2$, when logic constants $0,1$ are included in the set.}
\label{fig:visualization2}
\end{figure}

Table \ref{tab:2inputUniversalGate} depicts the number of logic that each gate can implement including  logic constants $0$ and $1$. Gates that are able to implement all $16$ outcomes are considered universal. Inputs of each gate are comprised of the gates that are self-generated, as well as $A, B$, $All 0(0)$ and $All 1(F)$. As an example, $NAND(7), NOR(1)$, A Imply B $(B)$, B Imply A $(D)$, A Nimply B $(4)$, and B Nimply A $(2)$ gates make up the coalescence of universal gates. This becomes implicit that by adding $0$ and $F$ gates as inputs to a gate increased both the number of possible outcomes and the universal gate count by $4$. Fig. \ref{fig:visualization2} shows that the number of logic functions created by a gate increases (red colored bars) when $0$ and $1$ constants are included in the set.
\subsection{Universal logic ($N=3$) including the logic constants 0, 1}
Table \ref{tab:3inputNonUniversalWithLgicFunction} shows the number of logic implemented by $31$ Non-Universal gates including the logic constants $0,1$. It is observed that a Non-Universal gate can create a maximum of $20$ logic functions out of $256$ possible gates. Including the logic constants in the set converted majority of the gates to the universal logic. The additional Non-Universal gates that became part of the Universal count after being a part of Set \{0, 1, G\} are shown in Table \ref{tab:3inputExtraUniversalGate}. For a  $3$ variable case, out of $256$ possible logic outcomes, $225$ are found Universal. Consequently, the ratio of Universal Gates to Non-Universal Gates increases dramatically as the number of inputs  increase.
\section{Decoding the class of a logic using its Multiplexer equivalent Circuits }
In this section, we will examine the multiplexer equivalent circuits of logic gates, which can be further used in determining the class of a logic. Consider a three-variable logic function in which one of the inputs can be assigned either $0$ or $1$. This can be done by connecting the variable to either $All0$ gate or an $AllF$ gate to switch between the first four and last four gates of the function. Due to the inclusion of $0$ and $1$ in the entire set, it can therefore function as a combination of two $2$-input gates, which can be selected by switching one of the variables. Now,let us consider a 3-variable function with inputs $A, B$ and $C$ to be implemented using a $2:1$ Multiplexer with $A$ as the select line. The input value $A$ can be utilised to switch between two distinct functions. It is observed that if any of the gates in the combination is  universal  that are used as inputs in the multiplexer equivalent circuit is a Universal gate, then the $3$-input gate also becomes a Universal gate (\textit{\textbf{Theorem 2}}). 
This is also true, if two gates in the multiplexer equivalent circuit forms a functionally complete set. They can be combined to create a function that can implement any function, resulting in a Universal logic (using \textit{\textbf{Property 1}}). Logic gates with $N$-inputs can be constructed using higher order multiplexers involving only $2$-input gates, the output of which are connected to input lines of the MUX. Thus, there are $(N-2)$ select lines in MUX which on receiving an input combination activates a specific minterm at the output.
\subsection{A Fast-Track approach for logic analysis using the Hexadecimal encoding of the gate}
A method has been described in this section whereby the hexadecimal representation of a logic gate can be used to infer its property without carrying out rigorous analysis on the logic of universality.
Thus, it is possible to deduce, based on the Multiplexer equivalent circuit  that all binary gates that consist of any of the six $2$-input Universal gates as input of the multiplexer equivalent circuit are Universal. This can also be validated by inspecting the hexadecimal encoding of gates. It is observed that for $N=2$ there are $6$ Universal Gates with hexadecimal values $1, 2, 4, 7, B,$ and $D$. 

\begin{table}[H]
\caption{Non-Universal gates for $N=3$ with the corresponding number of logic functions that they can implement}
\centering
\resizebox{0.75\linewidth}{!}{\begin{tabular}{|c|c|c|c|c|c|c|c|}
\hline
\textbf{Gate\#} & \textbf{Out} & \textbf{Gate\#} & \textbf{Out} & \textbf{Gate\#} & \textbf{Out} & \textbf{Gate\#} & \textbf{Out}\\\hline
00 & 1 & 0F & 8 & 33 & 8 & 3C & 16\\\hline
55 & 8 & 5A & 16 & 66 & 16 & 69 & 16\\\hline
80 & 9 & 88 & 9 & 96 & 16 & 99 & 16\\\hline 
A0 & 9 & A5 & 16 & A8 & 20 & AA & 5\\\hline
C0 & 9 & C3 & 16 & C8 & 20 & CC & 5\\\hline 
E0 & 20 & E8 & 20 & EA & 20 & EC & 20\\\hline
EE & 9 & F0 & 5 & F8 & 20 & FA & 9\\\hline 
FC & 9 & FE & 9 & FF & 1 & & \\\hline
\end{tabular}}
\label{tab:3inputNonUniversalWithLgicFunction}
\end{table}

\begin{table}[H]
\caption{List of extra Universal Gates including logic constants $0,1$}
\centering
\resizebox{0.97\linewidth}{!}{\begin{tabular}{|c|c|c|c|c|c|c|c|c|c|c|c|c|}
\hline
02 & 04 & 06 & 08 & 0A & 0C & 0E & 10 & 12 & 14 & 16 & 17 & 18 \\\hline
1A & 1C & 1E & 20 & 22 & 24 & 26 & 28 & 2A & 2B & 2C & 2E & 30 \\\hline
32 & 34 & 36 & 38 & 3A & 3E & 40 & 42 & 44 & 46 & 48 & 4A & 4C \\\hline
4D & 4E & 50 & 52 & 54 & 56 & 58 & 5C & 5E & 60 & 62 & 64 & 68 \\\hline
6A & 6C & 6E & 70 & 71 & 72 & 74 & 76 & 78 & 7A & 7C & 7E & 81 \\\hline
82 & 83 & 84 & 85 & 86 & 87 & 89 & 8A & 8B & 8C & 8D & 8E & 8F \\\hline
90 & 91 & 92 & 93 & 94 & 95 & 97 & 98 & 9A & 9B & 9C & 9D & 9E \\\hline
9F & A1 & A2 & A3 & A4 & A6 & A7 & A9 & AB & AC & AD & AE & AF \\\hline
B0 & B1 & B2 & B3 & B4 & B5 & B6 & B7 & B8 & B9 & BA & BB & BC \\\hline
BD & BE & BF & C1 & C2 & C4 & C5 & C6 & C7 & C9 & CA & CB & CD \\\hline
CE & CF & D0 & D1 & D2 & D3 & D4 & D5 & D6 & D7 & D8 & D9 & DA \\\hline
DB & DC & DD & DE & DF & E1 & E2 & E3 & E4 & E5 & E6 & E7 & E9 \\\hline
EB & ED & EF & F1 & F2 & F3 & F4 & F5 & F6 & F7 & F9 & FB & FD \\\hline
\end{tabular}}
\label{tab:3inputExtraUniversalGate}
\end{table}
 A logic gate in higher order variables ($N$$>$$2$) whose hexadecimal encoding contains any of digits from the given set of $6$ hexadecimal values becomes Universal. This is because the presence of any of these values in the final encoding implies that the multiplexer equivalent circuit uses the $2$-input universal gate in the input line. Thus, a logic whose encoding contains any of the $6$ digits; $1,2,4,7,B$ and $D$ are Universal. Therefore, from the hexadecimal encoding of the binary gates it can be ascertained that out of $256$ possible three-input gates, $(6\times16)+(10\times6)=156$ are Universal because all of them contain digits from the set $\{1, 2, 4, 7, B, D\}.$

It is implicit that none of the $31$ Non-Universal Gates have any of the $6$ hexadecimal digits included in their hexadecimal encoding, as represented in Table \ref{tab:3inputNonUniversalWithLgicFunction}. The remaining $69$ gates, however, are Universal despite the fact that they do not include any of the six gates from the set. This is because they  contain digits that make up a set that is functionally complete. This enables the identification of a group of logic connectives that can be combined to form a functionally complete set.

From the Hexadecimal encoding of 4-input binary gates, we can also deduce that $(225\times256)+(31\times225)=64575$ are Universal, as they contain the H-E of Universal Gates for $N=3$. The remaining $961$ gates can be  investigated in a similar way in  forming complete sets in order to locate additional Universal Gates. Without adding the complete set to construct Universal Gates, the ratio of Universal gates to total  gate count becomes $37.5\%$ for $N=2$, $87.89\%$ for $N=3$, and $98.53\%$ for $N=4$. Consequently, the increase in $N$ brings the count of Universal Gates closer to $100\%$ when logic constants  are included in the set. The class of a particular logic can be deduced using the predetermined set of conditions. Thus, a logic gate can be Universal on its own, Universal in a Complete Set that includes the constants $0$ and $1$, or entirely Non-Universal. The steps to classify a logic gate is illustrated using  case studies \ref{secgate85}, \ref{secgate46}, \ref{secgate4685}.

\subsection{3-Input Gate 85}
\label{secgate85}
Consider a $3$ input gate $85$, which expressed using $F$=$\overline{A}\overline{C}+ABC$. The truth table given in Table \ref{tab:Gate85}, suggests that the given function does not satisfy $F(1,1,1)=0$, making it a Non-Universal Gate on its own. However, given the set $\{0,85,FF\}$, the gate becomes Universal. The multiplexer equivalent circuit of gate $85$ shown in Fig. \ref{gate85} accepts two gates as inputs namely; NOT (Gate $5$) and AND (Gate $8$), that are employed in the  $0$ and $1$ input lines of the MUX respectively. By giving inputs to $A$ which is the select line of the MUX it is possible to switch between the two logic functions. Because  additional gates included in the set i.e., gate $0$ and $FF$ as well as the fact that \{AND, NOT\} forms a functionally complete set, the gate also behaves as a universal one.
\subsection{3-Input Gate 46}
\label{secgate46}
Consider a $3$ input gate $46$ which is given by, $F$=$\overline{A}.\overline{B}.C+B.\overline{C}$. The truth table given in Table \ref{tab:Gate46}, suggests that the given function does not satisfy $F(0,0,0)=1$, making it a Non-Universal Gate on its own. However, given the set $\{0,46,FF\}$, the gate becomes Universal. This is due to the fact that its Multiplexer equivalent consists of a Universal Gate i.e., Gate $4$ (Nimply). This is possible because of the additional gates in the set i.e., gate $0$ and $FF$. 
The multiplexer equivalent circuit of gate $46$ in Fig. \ref{gate46} accepts two gates namely; XOR (Gate $6$) and Nimply (Gate $4$), that are employed in the input lines 0 and 1 of the multiplexer, respectively. By switching select line between $1$ and $0$ it is possible to switch between these two different functions.
 
\begin{minipage}{0.468\textwidth}
\begin{minipage}{0.5\textwidth}
\begin{figure}[H]
\centering
\includegraphics[height=2cm]{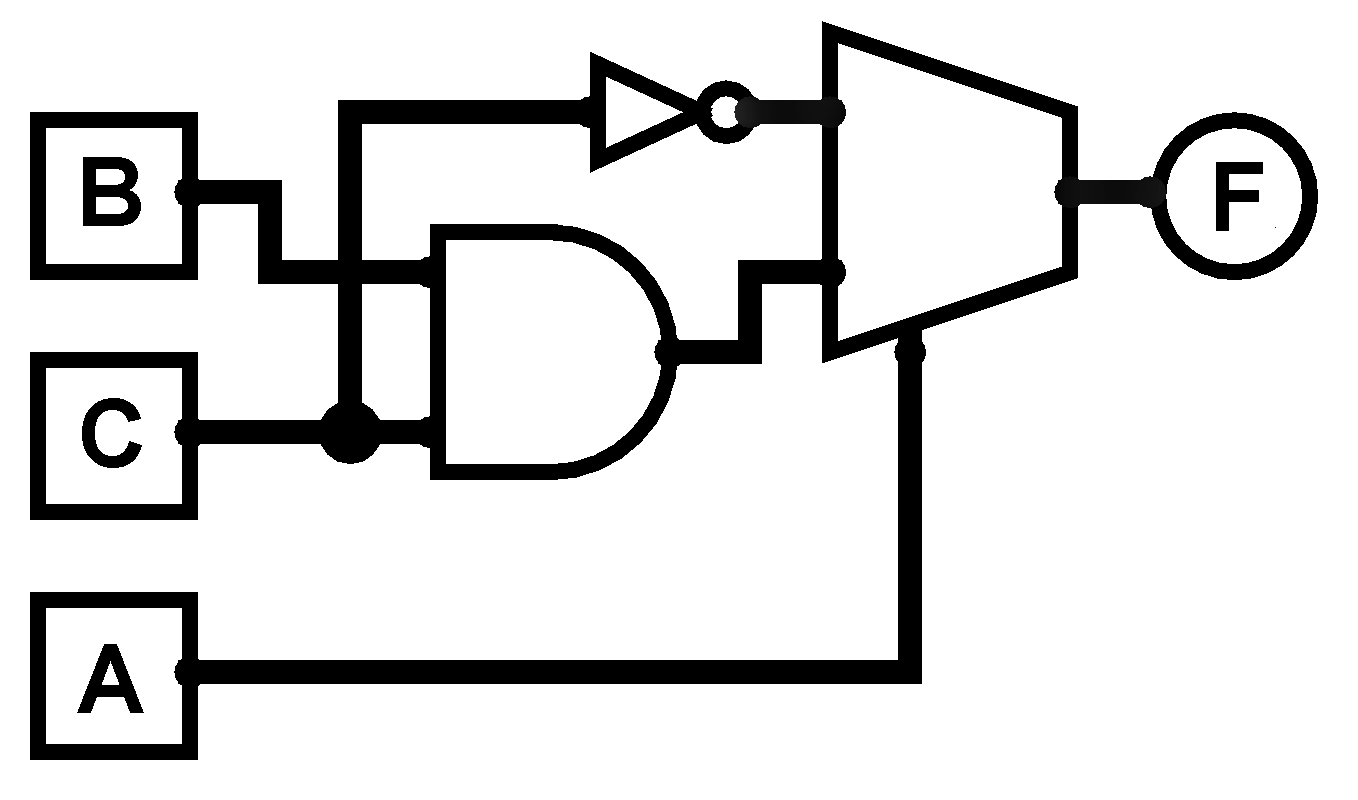}
\caption{Multiplexer equivalent circuit of Gate \#85}
\label{gate85}
\end{figure}
\end{minipage}
\begin{minipage}{0.3\textwidth}
\begin{table}[H]
\centering
\caption{Truth table of Gate \#85}
\resizebox{0.85\textwidth}{!}{\begin{tabular}{|c|c|c|c|c|c|c|c|c|c|c|}
\hline
\textbf{A} & \textbf{B} & \textbf{C} & \textbf{\#85} \\\hline
0 & 0 & 0 & 1 \\\hline
0 & 0 & 1 & 0 \\\hline
0 & 1 & 0 & 1 \\\hline
0 & 1 & 1 & 0 \\\hline
1 & 0 & 0 & 0 \\\hline
1 & 0 & 1 & 0 \\\hline
1 & 1 & 0 & 0 \\\hline
1 & 1 & 1 & 1 \\\hline
\end{tabular}}
\label{tab:Gate85}
\end{table}
\end{minipage}
\end{minipage}

\subsection{4-Input Gate 4685}
\label{secgate4685}
In a $4$ variable logic there are $65,536$ possible logic functions. Let us consider a  logic function with $4$-input variables   whose hexadecimal encoding is $4685$ and can be  expressed using the following switching function, 

 $F=\overline{A}.\overline{B}.\overline{D} + \overline{A}.B.C.D + A.\overline{B}.\overline{C}.D + A.C.\overline{D}.$

\begin{minipage}{0.468\textwidth}
\begin{minipage}{0.55\textwidth}
\begin{figure}[H]
\centering
\includegraphics[height=2.5cm]{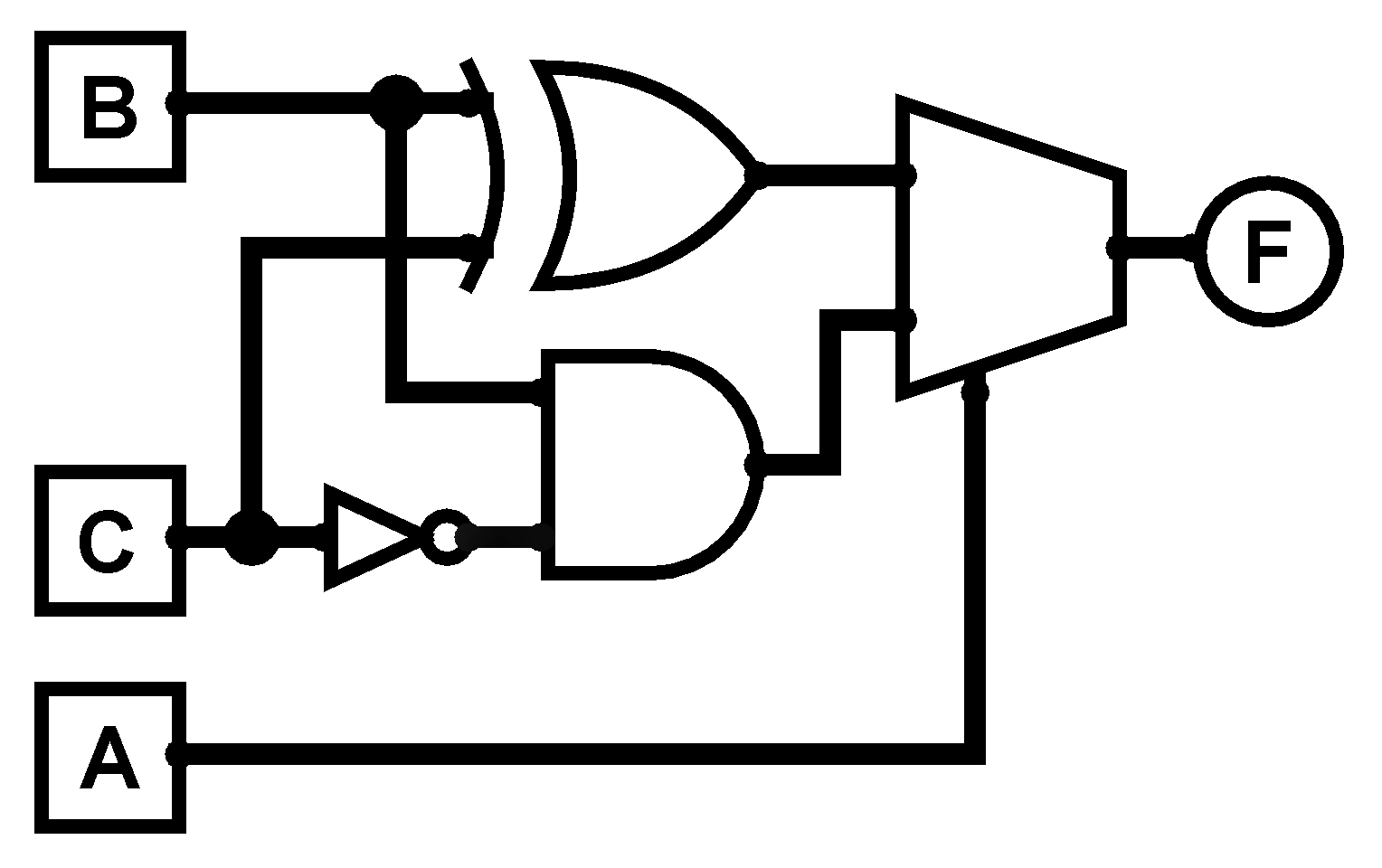}
\caption{Multiplexer equivalent circuit of Gate \#46}
\label{gate46}
\end{figure}
\end{minipage}
\begin{minipage}{0.3\textwidth}
\begin{table}[H]
\caption{Truth table of Gate \#46}
\centering
\begin{tabular}{|c|c|c|c|c|c|c|c|c|c|c|}
\hline
\textbf{A} & \textbf{B} & \textbf{C} & \textbf{\#46} \\\hline
0 & 0 & 0 & 0 \\\hline
0 & 0 & 1 & 1 \\\hline
0 & 1 & 0 & 1 \\\hline
0 & 1 & 1 & 0 \\\hline
1 & 0 & 0 & 0 \\\hline
1 & 0 & 1 & 0 \\\hline
1 & 1 & 0 & 1 \\\hline
1 & 1 & 1 & 0 \\\hline
\end{tabular}
\label{tab:Gate46}
\end{table}
\end{minipage}
\end{minipage}
\vspace{2mm}

\begin{figure}[!t]
\centering
\includegraphics[width=4.5cm]{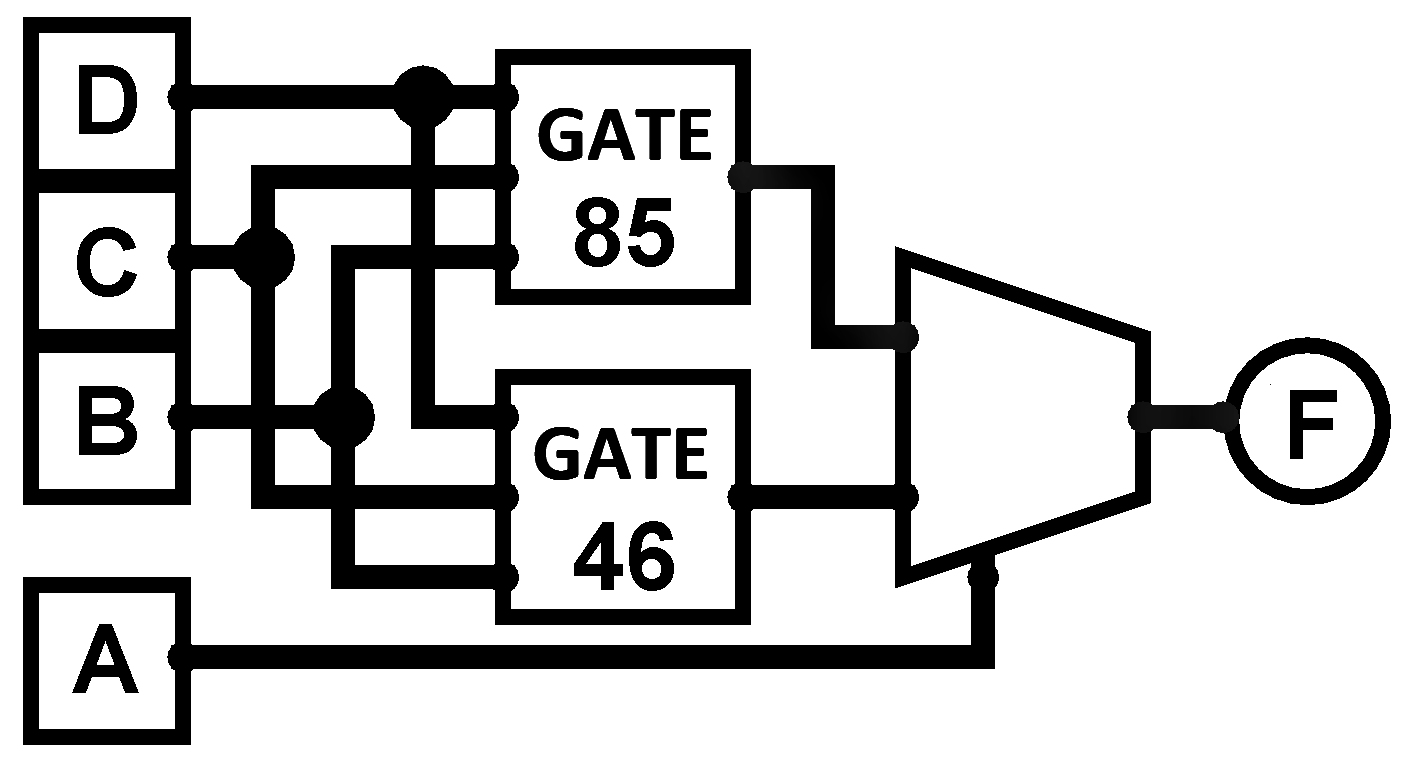}
\caption{2:1 MUX equivalent circuit of Gate \#4685}
\label{2gate4685}
\end{figure}

The truth table of gate $4685$ is given in Table. \ref{4685} shows that the given logic does not exhibit the property of self-dual functions. It is also observed that the gate satisfies both the conditions of $0$ and $1$ non-preserving functions, i.e.,
$F(0,0,0,0)=1$ and $F(1,1,1,1)=0$. Therefore, this is a Universal gate  on its own in compliance with the \textit{\textbf{Condition 4}}. 
If we introduce the complete set \{0000, 4685, FFFF\}, it is  observed that  the gate consists of gates $4$ ($Nimply$), $46$ and $85$, which are Non-Universal Gates by virtue but translated to universal logic when $0,1$ are included in the set. Multiplexer equivalent circuits of the logic gate using a $4:1$ MUX (considering $A$ and $B$ as the select lines) is shown in Fig. \ref{ta:gate4685}. When $A=0,B=0,$ multiplexer implements NOT gate (gate $5$), when $A=0,B=1,$ multiplexer implements AND gate (gate $8$), when $A=1,B=0,$ multiplexer implements XOR gate (gate $6$), when $A=1,B=1,$ multiplexer implements Nimply gate (gate $4$).
The same function implemented using
 $2:1$ MUX (considering $A$ as the select line)   is shown in Fig. \ref{2gate4685}.
A similar foregoing approach can be grabbed to derive the $2:1$ multiplexer equivalent circuit for the function using logic blocks consisting of gate $85$ and $46$. If we consider $D$ as the select input and $A,B,C$ as the inputs to the MUX, we can access  newly added sets of gates; gate $18$ (for $D=1$) and  $A3$ (for $D=0$). Thus, the H-E of the logic changes to $18A3$ when $D$  is the select input. However, the the condition of universality does not change, as the H-E consists of $1$ (NOR gate) that implies a universal logic. Changing the select line  alters the H-E of the accessible gates but does not change the overall nature of the  function. A quicker approach to analyse the property of a logic function is presented in \textbf{\textit{Algorithm $2$}}.
\begin{algorithm}[H]
\begin{algorithmic}[1]
\small
\caption{\textit{A faster approach to search Universal logic  gates}} 
\label{Algorithm:fastTrack}
\STATE \textbf{Input:} $Truth Table$ \textit{{//Truth Table of Function with n inputs.}}
\\\textbf{Output: }Universal or Non-Universal Gate
\STATE \textbf{Variable Initialization:}
\\$n$ = Number of Inputs
\\$M=2^n$ \textit{{//No. of input logic combinations.}}
\STATE {$TemplateGate[\;]=\{1,2,4,7,B,D\}$ //list of universal gates obtained from $n=2$ as shown in  Table I.}
\STATE {$|.|$ calculate the cardinality of the array}
\STATE Initialize $TT[\;]$ with the Values from Truth Table in hexadecimal
 \FOR  {$inputIndx=3; inputIndx<=n;inputIndx++$}
 \FOR  {$i=1; i<|TT|;i++$}

\STATE {$DigitIndx[]=GetDigitComb(TT[i]);$   \\//denoting each gate for n variable boolean logic ; the combination    would be till $n-1$ digit } 
\STATE $CountDigit=|DigitIndx|$
 \FOR  {$j=1; j<CountDigit;j++$}
 \FOR  {$k=1; k<|TemplateGate|;k++$}
 \IF {$DigitIndx[j]==|TemplateGate[k]|$}
      \STATE {
       $TemplateGate[\;].append=TT[i]$}
      \STATE break;
       \ENDIF
\ENDFOR
  
\ENDFOR
 \RETURN $TemplateGate$
\ENDFOR
\ENDFOR
\end{algorithmic}
\end{algorithm} 
\begin{minipage}{0.468\textwidth}
\begin{minipage}{0.5\textwidth}
\begin{figure}[H]
\centering
\includegraphics[width= 4cm]{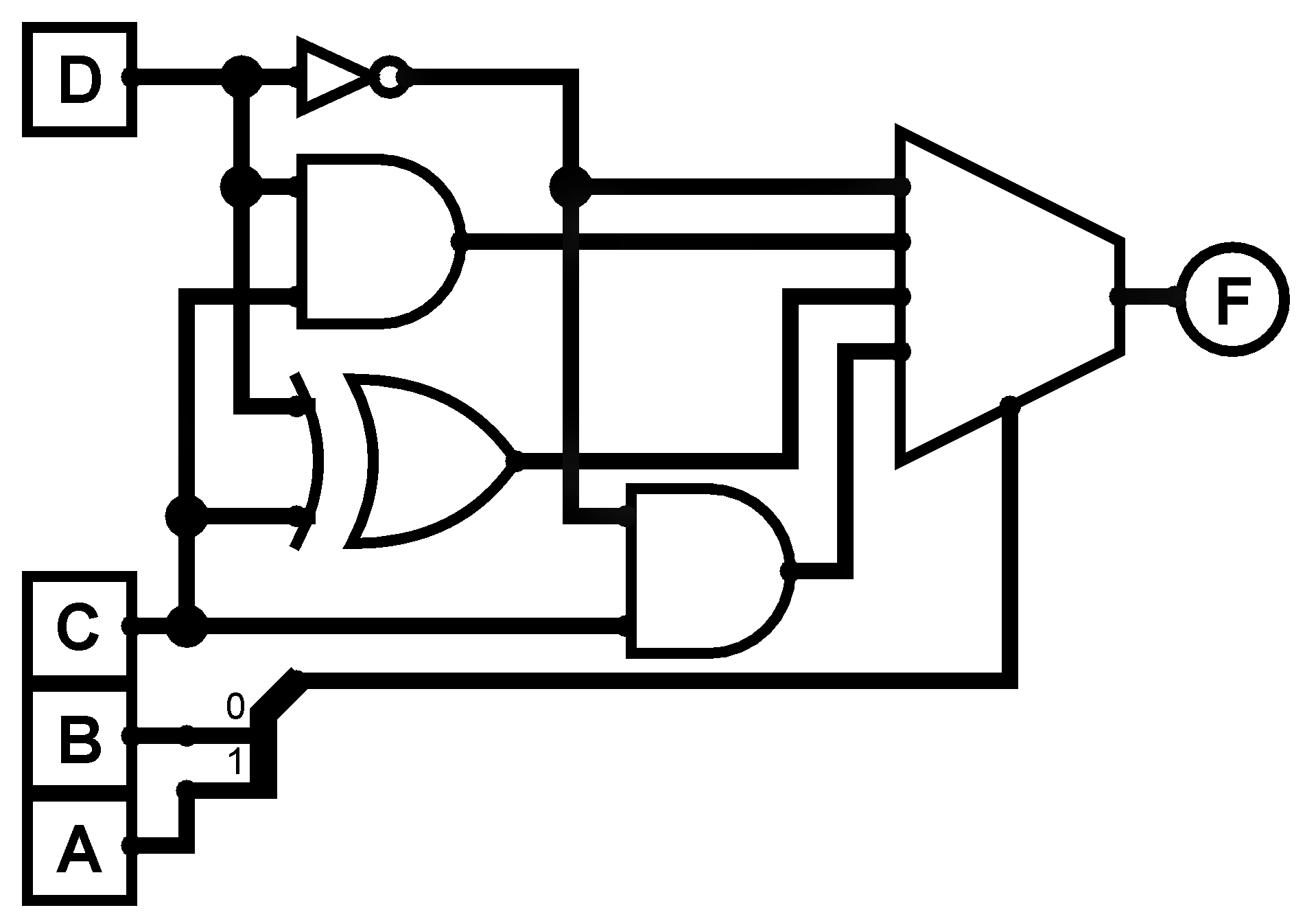}
\caption{4:1 MUX equivalent circuit of  universal logic \#4685}
\label{ta:gate4685}
\end{figure}
\end{minipage}
\begin{minipage}{0.35\textwidth}
\begin{table}[H]
\caption{Truth table of Universal Gate \#4685}
\centering
\resizebox{0.94\textwidth}{!}{
\begin{tabular}{|c|c|c|c|c|c|c|c|c|c|c|}
\hline
\textbf{A} & \textbf{B} & \textbf{C} & \textbf{D} & \textbf{\#4685} \\\hline
0 & 0 & 0 & 0 & 1 \\\hline
0 & 0 & 0 & 1 & 0 \\\hline
0 & 0 & 1 & 0 & 1 \\\hline
0 & 0 & 1 & 1 & 0 \\\hline
0 & 1 & 0 & 0 & 0 \\\hline
0 & 1 & 0 & 1 & 0 \\\hline
0 & 1 & 1 & 0 & 0 \\\hline
0 & 1 & 1 & 1 & 1 \\\hline
1 & 0 & 0 & 0 & 0 \\\hline
1 & 0 & 0 & 1 & 1 \\\hline
1 & 0 & 1 & 0 & 1 \\\hline
1 & 0 & 1 & 1 & 0 \\\hline
1 & 1 & 0 & 0 & 0 \\\hline
1 & 1 & 0 & 1 & 0 \\\hline
1 & 1 & 1 & 0 & 1 \\\hline
1 & 1 & 1 & 1 & 0 \\\hline
\end{tabular}}
\label{4685}
\end{table}
\end{minipage}
\end{minipage}

\section{Applicability of the Universal Logic Gates to certain applications: A Case Study}
NAND and NOR gates are the most fundamental gates that can be created with the least amount of hardware in CMOS technology. These versatile universal logic gates, which can execute a large number of different logic functions, are playing an increasingly critical role in modern digital systems. As a result, all Boolean Functions are implemented using NAND and NOR equivalent circuits. The amount of hardware needed to generate a function increases if other Universal Gates are employed in CMOS implementation. Whereas, the most fundamental units for TTL-based implementation are AND, OR, and NOT gates; hence, these gates are utilised to construct functions to minimise hardware complexity in a technology.

\begin{figure}[!t]
\centering
\includegraphics[height=1.82cm]{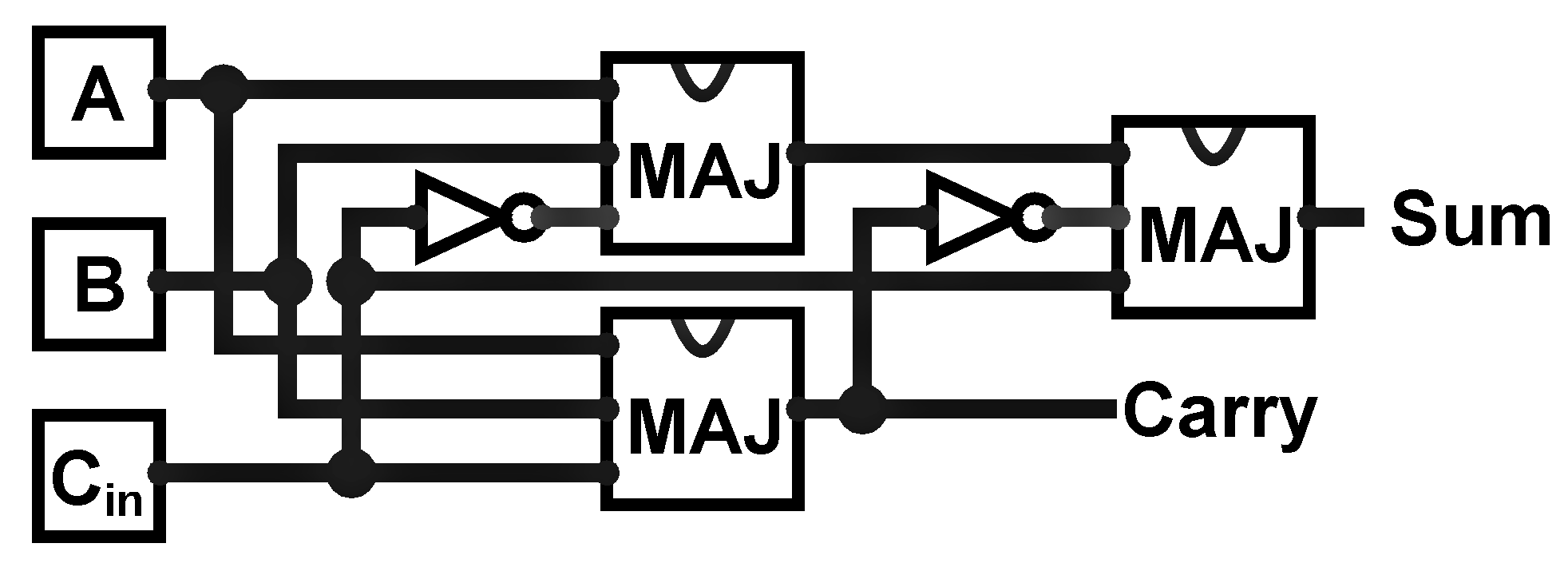}
\includegraphics[height=1.91cm]{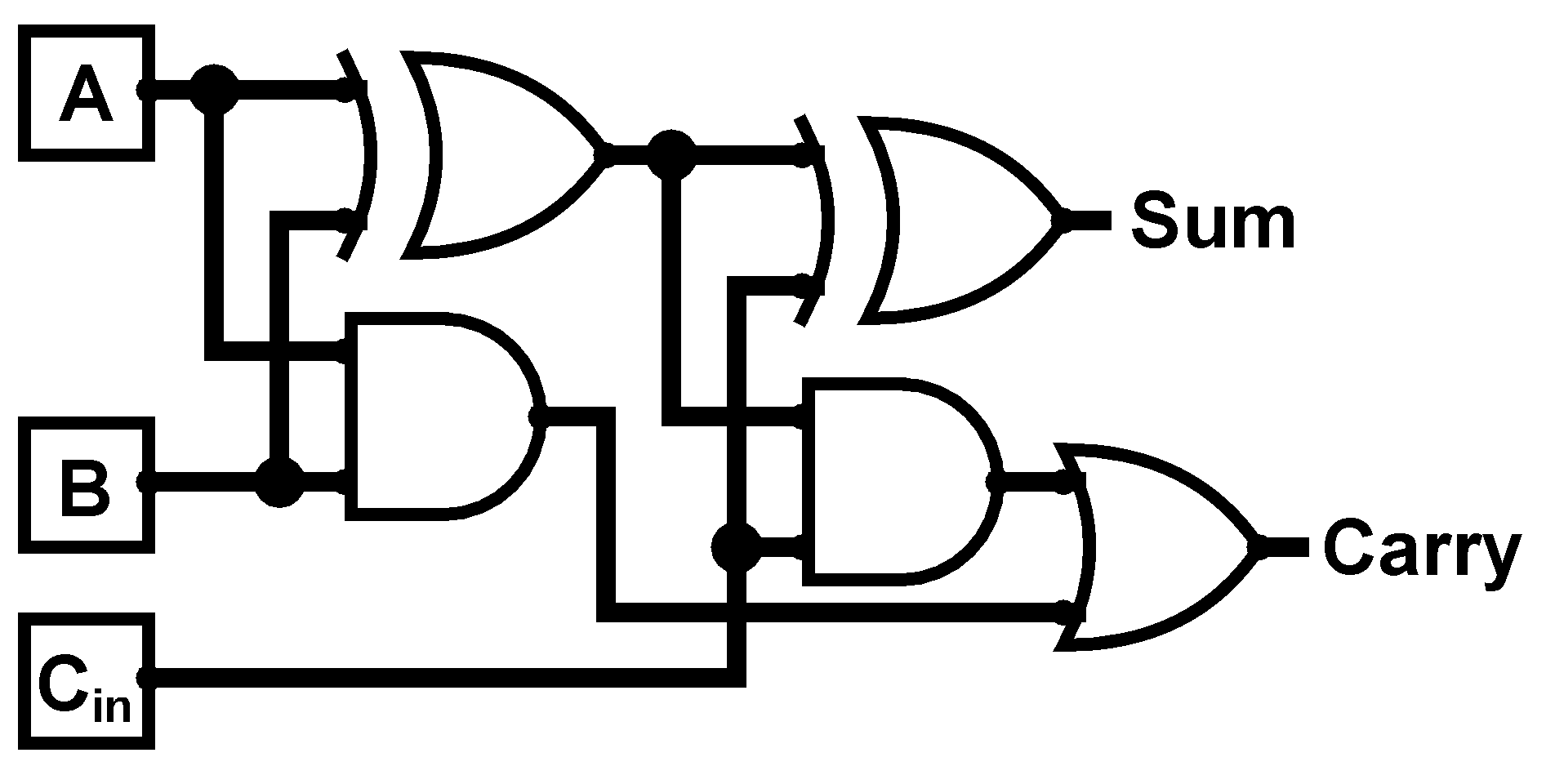}
\caption{Full Adder Circuit implemented using a) Majority-Inverter (LHS) logic b) TTL logic (RHS)}
\label{AIG}
\end{figure}

\begin{figure}[!t]
\centering
\includegraphics[width=0.35\textwidth]{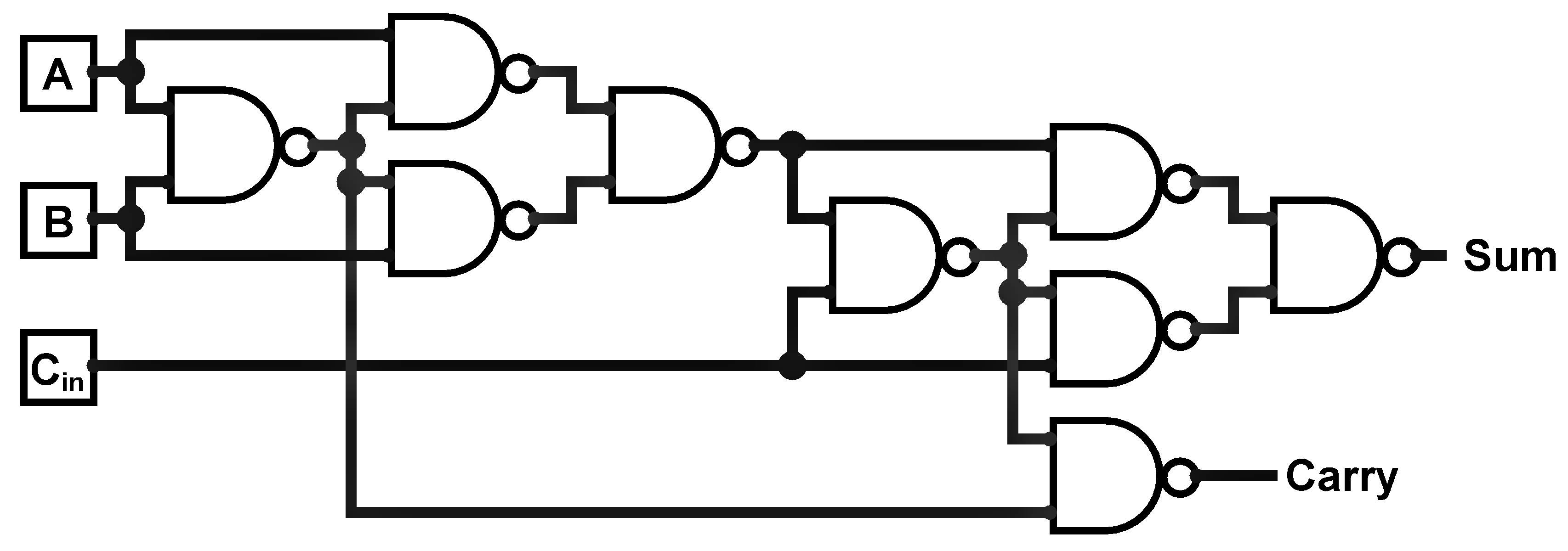}
\caption{Full Adder implemented using only NAND Logic}
\label{ADD}
\end{figure}

Low-cost hardware solutions are generally technology-dependent, relying on certain technologies or components to achieve cost-effectiveness. In  applications like Quantum Dot Cellular Automata (QCA) \cite{7538997,Farrelly2020reviewofquantum}, Majority Gate and NOT Gate are the most fundamental logic units. Majority gate with H-E $E8$ is represented by $f=AB+BC+CA$. Employing Majority gate instead of traditional logic gates it is possible to reduce both the number of connections between gates and the total number of gates. For example, to create a Full Adder circuit, Majority and Inverter gates based design shown in Fig. \ref{AIG}a) requires $5$ logic Gates (3 Majority and 2 Inverter). Only three gates, including two inverters, are needed to construct a full adder.  A Boolean implementation using NAND and NOR gates, on the other hand, necessitates a more complex design with 9 gates (requiring 36 transistors). Thus,  Majority-Inverter logic based approach is more hardware efficient than NAND and NOR implementations of the same.
A vast majority of other logic  functions can be efficiently implemented using majority logic.  The other two architectures using only NAND logic and TTL logic are shown in Fig. \ref{ADD} and \ref{AIG}b respectively.  
In essence, the technologies utilised to tackle a problem are not necessarily application independent; rather, attaining a hardware efficient solution is highly dependent on the application domain.
As a result, certain basic gates are widely used in implementation technologies in order to reduce the overall hardware needs of the circuit. 

\section{Conclusion}

In this work, a detailed and diverse study on the property of universality of a logic is carried out. Further, the number of Universal logic gates for a given number of input variables is determined. It is observed that there $56$ Universal gates for $N=3$ and the ratio of Universal Gates to all possible logic gates for $N>3$ saturates to $\frac{1}{4}$ or 0.25. We have also discussed how the properties of Universal Gates alter when a complete set $\{0, 1, G\}$ is considered. Introducing the Multiplexer equivalent property of a logic, the reasoning behind the universality of a complete set is demonstrated. The number of universal logic is increased from $2$ to $6$ for $N=2$ and $56$ to $225$ for $N=3$. The increased availability of Universal Gates in the future makes such data extremely useful for logic synthesis of a specified function. By deducing the logic for minimising a function with the included set of Universal gates, one can go on to design cheaper arithmetic circuits. The applicability of the newly found universal logic has been discussed in great detail with the aim to reduce the hardware in specific applications. The presented logic analysis can be extrapolated to larger base number systems to locate the Universal gates. In future, we move forward for further logic minimization methods employing these gates. Also, we will try to explore the characteristics of the newly found universal logic gates and investigate their potential uses with respect to logic design and synthesis in certain fields of study. 



\begin{thebibliography}{1}

\bibitem{4468044}Zhang, W. \& Wu, N. Smart Universal Multiple-Valued Logic Gates by Transferring Single Electrons. {\em IEEE Transactions On Nanotechnology}. \textbf{7}, 440-450 (2008)

\bibitem{9864039}Jiao, S., Feng, J., Zhang, L., Wu, D. \& Shen, Y. Optical Logic Gate Operations With Single-Pixel Imaging. {\em IEEE Journal Of Selected Topics In Quantum Electronics}. \textbf{29}, 1-8 (2023)
\bibitem{9951401}Kolay, S., Bandyopadhyay, M. \& Chattopadhyay, S. Design and Testing of Digital Logic Gates Using HCS Macro-Model. {\em IEEE Transactions On Circuits And Systems II: Express Briefs}. \textbf{70}, 1134-1138 (2023)
\bibitem{9762783}Charlot, N. \& Gauthier, D. Sensitivity of a Chaotic Logic Gate. {\em IEEE Transactions On Circuits And Systems II: Express Briefs}. \textbf{69}, 3339-3343 (2022)
\bibitem{631214}Lin, C. \& Marek-Sadowska, M. On designing universal logic blocks and their application to FPGA design. {\em IEEE Transactions On Computer-Aided Design Of Integrated Circuits And Systems}. \textbf{16}, 519-527 (1997)
\bibitem{Liu2021-ih}Liu, L., Liu, P., Ga, L. \& Ai, J. Advances in applications of molecular logic gates. {\em ACS Omega}. \textbf{6}, 30189-30204 (2021,11)
\bibitem{Noiri2022}Noiri, A., Takeda, K., Nakajima, T., Kobayashi, T., Sammak, A., Scappucci, G. \& Tarucha, S. Fast universal quantum gate above the fault-tolerance threshold in silicon. {\em Nature}. \textbf{601}, 338-342 (2022,1), https://doi.org/10.1038/s41586-021-04182-y

\bibitem{803116}Lima, F., Johann, M., Guntzel, J., Carro, L. \& Reis, R. A tool for analysis of universal logic gates functionality. {\em Proceedings. XII Symposium On Integrated Circuits And Systems Design (Cat. No.PR00387)}. pp. 184-187 (1999)
\bibitem{AC}Church, A. Introduction to Mathematical Logic. (Princeton University Press,1956)
\bibitem{AFPM}Friedman, A. \& Menon, P. Theory and Design of Switching Circuits. (Computer Science Press, Woodland Hills,1975)

\bibitem{mitra2023low}Mitra, S., Goenka, A. \& Naskar, M. A Low Latency and Compact GCD Design using an Intelligent Seed-Selection Scheme of LL-PRNG. {\em IEEE Transactions On Computer-Aided Design Of Integrated Circuits And Systems}. (2023)
\bibitem{GK}Klir, G. Introduction to the Methodology of Switching Circuits. (van Nostrand, New York,1972)
\bibitem{AM}Mukhopadhay, A. Complete sets of logic primitives. (Academic Press, New York,1971)
\bibitem{WJ}Jevons, W. Pure Logic. (London,1864)
\bibitem{EP}Post, E. The Two-Valued Iterative Systems of Mathematical Logic. (Princeton University Press, Princeton, NJ,1941)
\bibitem{7538997}Reis, D., Campos, C., Soares, T., Neto, O. \& Torres, F. A Methodology for Standard Cell Design for QCA. {\em 2016 IEEE International Symposium On Circuits And Systems (ISCAS)}. pp. 2114-2117 (2016)
\bibitem{Farrelly2020reviewofquantum}Farrelly, T. A review of Quantum Cellular Automata. {\em Quantum}. \textbf{4} pp. 368 (2020,11), https://doi.org/10.22331/q-2020-11-30-368

\end{thebibliography}

\end{document}